\documentclass[preprint,showpacs,preprintnumbers,amsmath,amssymb]{revtex4}

\usepackage{graphicx}
\usepackage{dcolumn}
\usepackage{bm}

\begin{document}

\title{Voltage rectification effects in mesoscopic superconducting triangles: experiment and modelling}

\author{N. Schildermans$^1$, A.B. Kolton$^2$, R. Salenbien$^1$, V.
I. Marconi$^2$, A. V. Silhanek$^1$, V. V. Moshchalkov$^1$}
\affiliation{$^1$INPAC - Institute for Nanoscale Physics and
Chemistry, Nanoscale Superconductivity and Magnetism \& Pulsed
Fields Group, K.U.Leuven, Celestijnenlaan 200D, B-3001 Leuven,
Belgium.\\
$^2$Dep. de F{\'i}sica At\'omica, Molecular y Nuclear,  Universidad Complutense de Madrid, 28040 Madrid, Spain.}

\date{\today}
\begin{abstract}
The interaction of externally applied currents with persistent
currents induced by magnetic field in a mesoscopic triangle is
investigated. As a consequence of the superposition of these
currents, clear voltage rectification effects are observed. We
demonstrate that the amplitude of the rectified signal strongly
depends on the configurations of the current leads with the lowest
signal obtained when the contacts are aligned along a median of
the triangle. When the contacts are aligned off-centered compared
to the geometrical center, the voltage response shows oscillations
as a function of the applied field, whose sign can be controlled
by shifting the contacts. These results are in full agreement with
theoretical predictions for an analogous system consisting of a
closed loop with a finite number of identical Josephson junctions.

\end{abstract}

\pacs{74.25.Fy,74.25.Sv,74.78.Na,73.40.Ei, 74.81.Fa, 85.25.Cp}

\maketitle

A considerable attention has been attracted in the last years to
the mechanisms responsible for the ratchet effects in a broad
variety of physical systems such as colloids\cite{libal06},
granular materials\cite{farkaz99}, fluids\cite{matthias03}, atoms
in optical traps\cite{gommers06}, electrons in semiconductor
heterostructures\cite{linke03} and Josephson
systems\cite{zapata96,falo99,shalom05}. In all cases a net flux of
particles driven by a zero average alternating excitation results
from the interaction of the media with an asymmetric potential.
This behavior has been also theoretically predicted and
experimentally corroborated for the motion of quantum flux lines
in superconducting samples with a pinning landscape lacking
inversion symmetry \cite{lu07, cole06, desouzasilva06,
vandevondel05, Wambaugh, villegas03, silhanek07}. In
superconducting systems, this effect manifests itself as a
non-zero dc voltage even when an ac excition is sent through the
superconductor, thus acting as a voltage rectified. Interestingly
it has been recently demonstrated that the presence of this
voltage rectification in a superconductor do not necessarily imply
the motion of vortices in an asymmetric pinning potential but
might also result from non symmetric current distributions in the
superconducting sample \cite{joris}.

Indeed, first Dubonos {\it et.al.}\cite{dubonos03} reported
rectification effects in asymmetric superconducting rings. Later
on, Morelle {\it et. al.}\cite{morelle04} showed that similar
effects are observed in singly connected structures if the current
injection is off-centered. In both cases the effect was attributed
to an asymmetry in compensation or reinforcement of an external
bias current by the field induced persistent currents, causing a
difference in critical current for a positive or negative applied
external current. More recently Van de Vondel {\it et.al.}
\cite{joris} showed that both kinds of rectification, due to
ratchet vortex motion and due to current compensation effects can
coexist in superconducting samples with periodic arrays of
triangular antidots.

In this work we investigate the influence of the position of the
current/voltage probes on the resultant rectification effect in
microsized superconducting triangles.
We show that an ac current injected above the geometrical center
of the triangle gives an opposite rectification signal than for
current injection below the geometrical center of the triangle. In
addition we show that a lower signal is obtained if the contacts
are attached along a median of the triangle, so that upper and
lower part of the triangle are symmetric around this line.
This result demonstrates that the superposition of a field induced
persistent current with the bias current qualitatively accounts
for the observed phenomena. To interpret these data, we also used
a theoretical model system consisting of a closed loop of $N$
Josephson junctions containing the necessary ingredients
(persistent and bias currents) to reproduce the experimental
findings.

\begin{figure}
\begin{center}
\includegraphics[width=0.5\linewidth]{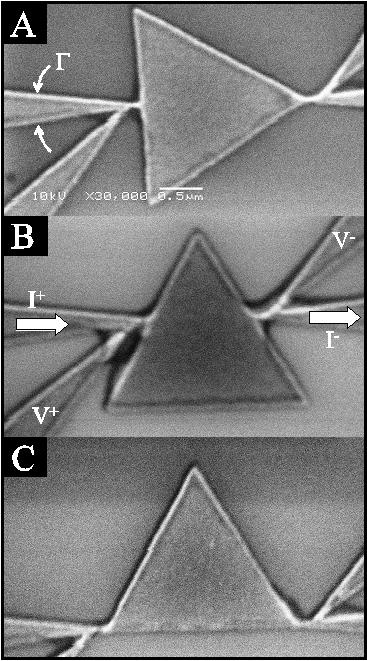}
\end{center}
\caption{Scanning electron microscopy image of the superconducting
equilateral Al triangle of 2.25 $\mu$m side length with wedge
shaped current and voltage contacts with an opening angle
$\Gamma=15^\circ$. }\label{fig:SEM}
\end{figure}

The superconducting triangles are made of a 50-nm-thick Al film
thermally evaporated on Si/SiO$_2$. The structures are obtained by
deposition through an e-beam patterned resist mask followed by a
lift-off procedure. All samples consist of an equilateral triangle
with an area of $S=2.2~\mu$m$^2$ and four wedge-shaped
current/voltage contacts. The typical superconducting coherence
length estimated from unpatterned films is about $\xi$(0)$=120$
nm. Three different contacts configurations are investigated:
current injected along a median (see the scanning electron
microscopy (SEM) image in Fig.~\ref{fig:SEM}(a)), current injected
above the geometrical center (Fig.~\ref{fig:SEM}(b)), and current
injected at the base of the triangle (Fig.~\ref{fig:SEM}(c)). From
hereon we refer to these samples as sample A, sample B, and sample
C, respectively.

\begin{figure}
\begin{center}
\includegraphics[width=1\linewidth]{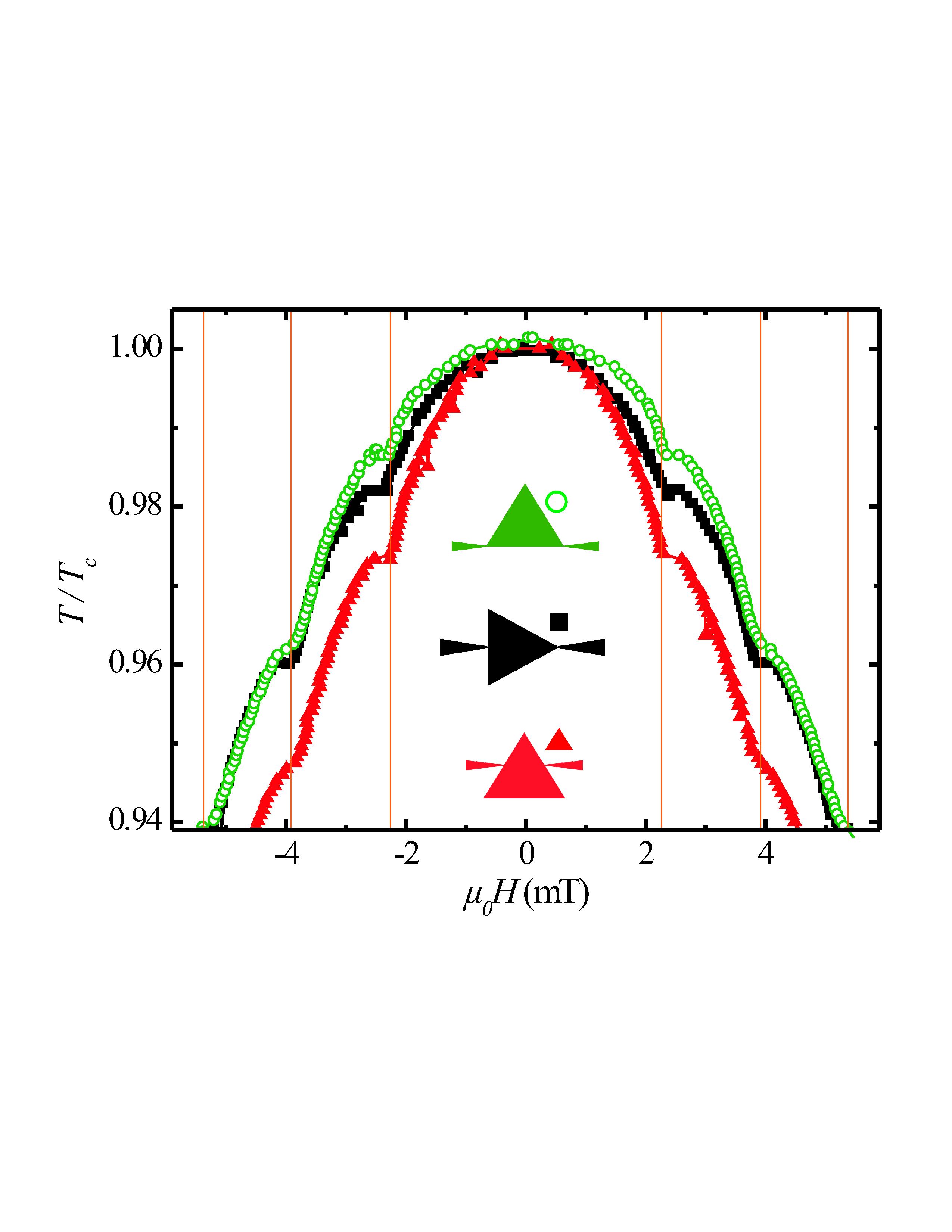}
\end{center}
\caption{(color online) Superconductor/Normal metal phase
boundaries determined by a 10\% criterium of the normal state
resistance. The vertical lines indicates the theoretical expected
field values for the Little-Parks oscillations $L \rightarrow L+1$
in a triangle with a surface of $S=2.05~\mu$m$^2$.} \label{fig:PB}
\end{figure}

The phase boundary for each of the studied samples is summarized
in Fig.~\ref{fig:PB} using an ac drive of 0.1 $\mu A$
peak-to-peak. Here $T_c(H)$ is estimated by a resistance criterion
of 10 percent of the normal state value. The obtained critical
temperatures $T_{c}(0)$ for samples A, B and C are 1.365 K, 1.355
K and 1.34 K, respectively. All phase boundaries exhibit clear
Little-Parks oscillations with local minima at fields $H_L$ where
the vorticity switches from $L$ to $L+1$\cite{tinkbook}. The
vertical lines in Fig.~\ref{fig:PB} show theoretical estimates of
the geometry-dependent transition fields $H_L$ for a triangular
sample of area $S=2.05~\mu$m$^2$\cite{chibotaru01}. This value is
in a good agreement (within 7\%) with the area estimated from the
SEM images shown in Fig.~\ref{fig:SEM}. The most obvious feature
in Fig.~\ref{fig:PB} is the different field dependence of $T_c$
for the three studied samples with the higher $T_c(H)$ for sample
C and lower $T_c(H)$ for sample B. Since the only difference
between samples is the position of the contact leads, the observed
discrepancy in $T_c(H)$ can be unambiguously attributed to the
influence of these contacts on the nucleation of the
superconducting condensate. It is a well established
fact\cite{fomin97ssc, morelle02, schweigert} that surface
superconductivity is greatly enhanced in wedge-shaped structures.
Since the contact leads have a narrower apex angle (15$^\circ$)
than the triangle (60$^\circ$), superconductivity starts to appear
at these points first. Under these circumstances the transition to
the superconducting state is expected to occur at higher
temperatures in sample C where both wedges, the contacts and the
vertices of the triangle, reinforce the surface nucleation effect.
Similar effects were anticipated by de Gennes and Alexander for
small superconductor with leads attached to
it\cite{degennes:81arm, Alex83}.

\begin{figure}
\begin{center}
\includegraphics[width=0.5\linewidth]{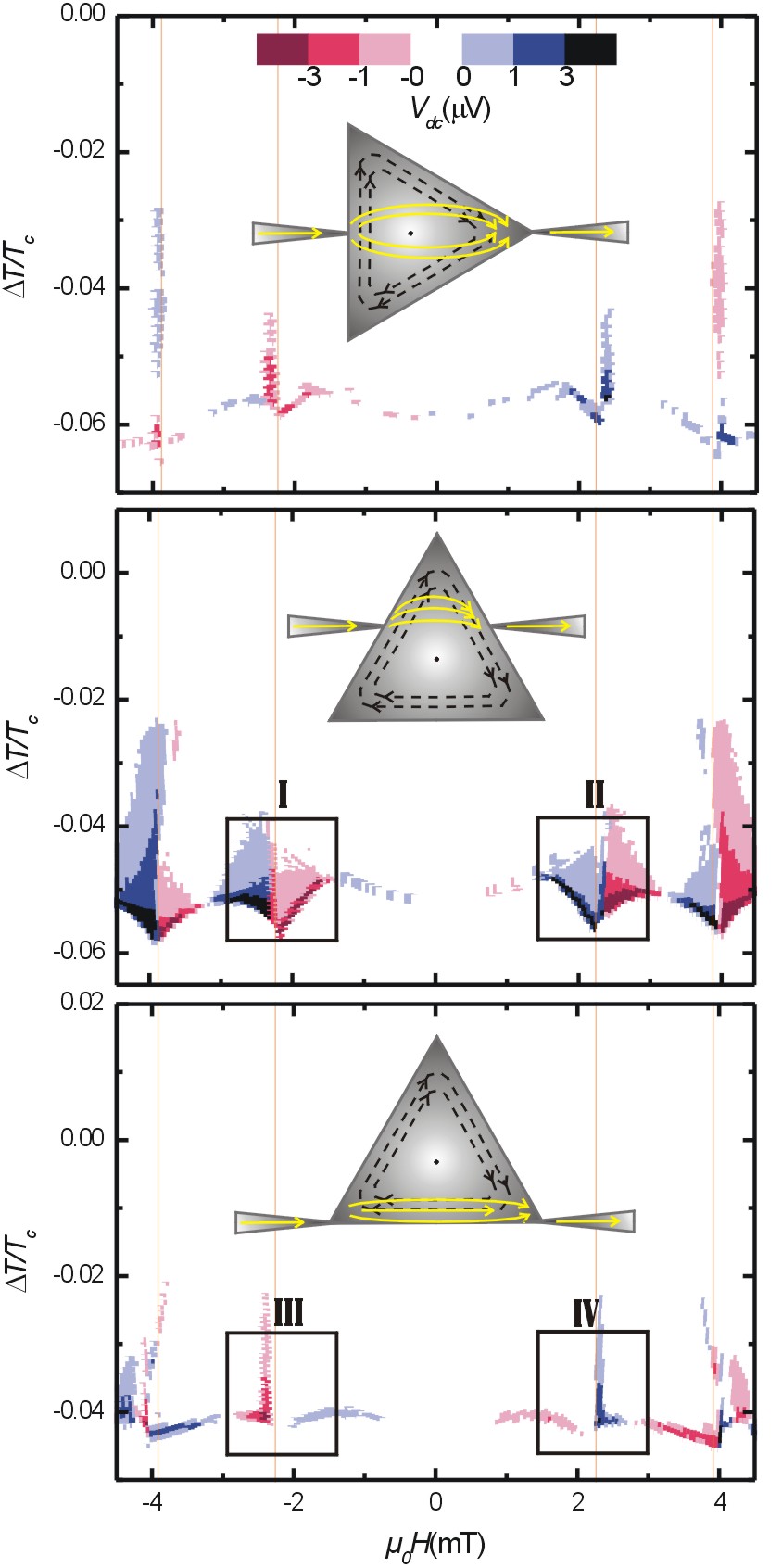}
\end{center}
\caption{(color online) Rectification signal obtained with an ac
excitation of 10 $\mu$A and frequency of $\sim$3.4 kHz as a
function of field and temperature for the sample A (upper panel),
B (middle panel), and C (lower panel). The vertical lines indicate
the theoretically expected field values for the Little-Parks
oscillations. The oscillating dc voltage is presented in a color
scale, from positive (blue) to negative (red). The data are
presented here with a parabolic background subtracted, so
$\Delta$T = $T_c$(H)-($T_c$(0) - b H$^2$), with b a constant
different for each sample. The inset in each panel gives a
schematic drawing of the circulating persistent current (black)
and the applied current (yellow) for that contact
configuration.}\label{fig:ratchet}
\end{figure}

Let us now focus on the interaction of the externally applied
currents with the persistent circular currents induced by the
magnetic field. In order to resolve better the difference between
positive and negative bias current we apply an ac drive with
amplitude of 10 $\mu$A peak-to-peak and frequency 3837 Hz while
measuring the average dc voltage. In this way, when the samples
are in the normal state ($T>T_c$) or in absence of screening
currents ($H=0$), the dc output voltage should be zero, as indeed
observed. In contrast to that, if screening currents are present,
as a result of an applied non zero homogeneous field, then the
superposition of the applied current with the circulating
persistent currents in the triangle gives a different contribution
when they reinforce than when they counteract each other. In other
words, a net dc voltage signal is recorded. Fig.~\ref{fig:ratchet}
shows the measured dc voltage $V_{dc}$ as a function of the
applied magnetic field and temperature $\Delta$T$/T_c$ for the
samples A, B, and C. The data are here presented with a parabolic
background subtracted, so $\Delta$T = $T_c$(H)-($T_c(0)$ - b
H$^2$), with b a constant different for each sample. In order to
make a reliable comparison of the measured signal between
different samples we have normalized $V_{dc}$ by the distance
between the voltage contacts.

Sample B has the same contact configuration as the one earlier
reported in ref\cite{morelle04}, with the leads placed above the
geometrical center of the triangle. As expected, this sample
reproduces the previous results, namely an abrupt change of sign
in the dc response (see the color changes in frames I and II in
Fig.~\ref{fig:ratchet}) every time the vorticity of the system
changes from $L$ to $L+1$, which is associated with the reversal
of the persistent currents, and a smooth crossover
in between two consecutive $H_L$ fields associated with the
progressive reduction and later inversion of the screening
currents.
The origin of the rectified voltage is related to the unbalanced
distribution of the external applied current and their
compensation with the persistent currents circulating around the
geometrical center. Since the current leads are located above the
geometrical center, the applied current mainly flows trough the
upper part of the triangle, causing an asymmetry by compensating
(or reinforcing) the screening currents more in the upper part
then in the lower part. A schematic drawing of the circulating
persistent current and the applied current are shown as an inset
in each graph of Fig.~\ref{fig:ratchet}.

Notice that besides the above mentioned voltage sign reversals
related to Little-Parks oscillations, an unexpected sign reversal
is observed in the Meissner phase for sample B (middle panel in
Fig.~\ref{fig:ratchet}). The fact that this effect, unlike the
Little-Parks oscillations, is much weaker and not systematically
observed for all measured samples indicates that it is sample
dependent and cannot be related to the circulating screening
currents.

According to the above described scenario, if the line along which
the current is inserted is not shifted compared to the center of
the circulating persistent currents, as in sample A, no
rectification effects should be observed. Since the upper part of
the triangle is in this case a mirror image of the lower part, the
current is distributed equally around the center of the triangle,
and no asymmetry in compensation ( or reinforcement) is expected.
This is consistent with the strongly reduced signal detected in
sample A (about three times weaker in amplitude and located in a
smaller temperature-field area), in comparison with sample B. The
origin of this small signal likely lies in the inevitable minor
asymmetries produced by shadow effects during the material
deposition.

The most compelling evidence that indeed the observed
rectification effects originate from the direct superposition of
the external current $I_{appl}$ and the field induced persistent
circular currents $I_{per}$ comes from the measurements shown in
the lower panel of Fig.~\ref{fig:ratchet} corresponding to sample
C. In sample C, unlike sample B, the current injection is well
below the geometrical center of the triangle and therefore the
situation should be reversed in comparison to sample B. In other
words, for the fields where a positive signal is recorded in
sample B (indicating that the positive current is reinforced in
the top of the triangle) an opposite sign is expected for sample C
(i.e. the positive current is compensated at the base of the
triangle). This reversal between samples B and C is clearly
visible for vorticity $L = 1$ by comparing the indicated sections
I, II with III, IV in Fig.~\ref{fig:ratchet}.

\begin{figure}
\begin{center}
\includegraphics[width=0.9\linewidth]{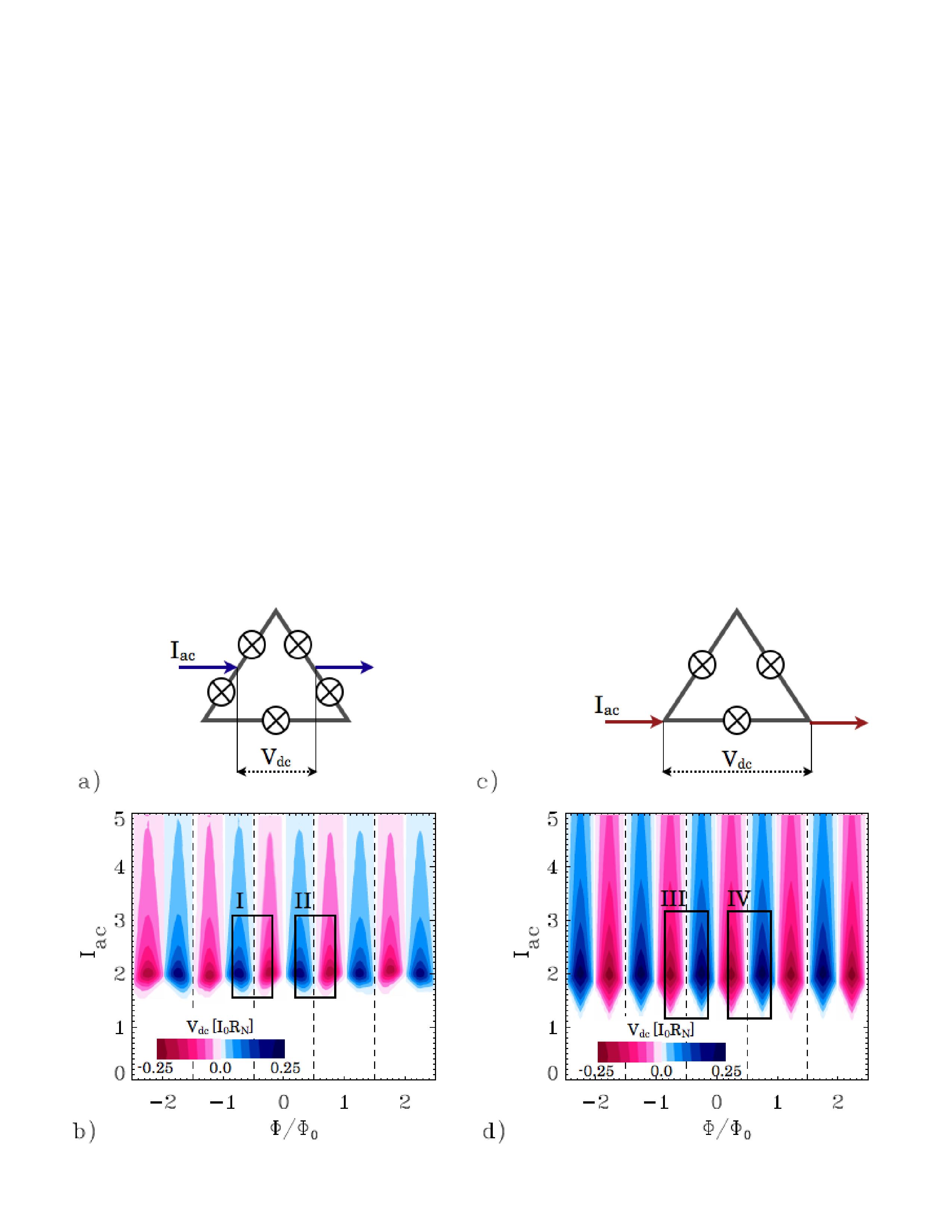}
\end{center}
\caption{(color online) Schematic drawing  of the superconducting
triangle viewed either as a ring of (a) $N=5$ Josephson junctions
(see Fig.1, sample B)  or  (c) $N=3$ JJ (see Fig.1, sample C),
depending on the contact positions.
 Below, corresponding contour plots of the rectified voltage, $V_{dc}$,  as
function of magnetic flux, $\Phi/\Phi_0$, and amplitude of the
applied ac-current, $I_{ac}$, for $N=5$ (b) and  $N=3$ (d). Note
the voltage sign difference by comparing  equivalent sections I,II
vs III,IV.}\label{fig:JJM}
\end{figure}

As we already briefly mentioned above, the necessary ingredients
to observe the sort of voltage rectification described here are
(i) field induced persistent currents and (ii) off-center
injection of external currents. This recipe suggests that similar
rectification effects should be also present in every system with
a persistent current and an asymmetric current path, thus
resulting in a difference in critical current for a positive or
negative applied current. An example of such a system fulfilling
these conditions consists of a closed loop with a number $N>2$ of
identical Josephson junctions (JJ). A complementary model with
$N=2$ and unequal JJ has been analyzed recently by
Berger\cite{berger04}. Without losing generality the main effects
can be seen in two simple configurations: a ring with three
junctions and one with five junctions, as schematically depicted
in Fig.~\ref{fig:JJM}.

The main assumptions for the calculations are that the
superconducting order parameter $\rho({\bf r})=|\rho({\bf
r})|e^{i\theta({\bf r})}$ is such that $|\rho({\bf r})|=\rho_0$
with $\rho_0$ the same constant on all islands, $\theta({\bf r})$
is spatially constant in each island, and that the weak links
between them can be modeled as identical SNS junctions. It is also
assumed that the total magnetic field $B$ is spatially and
temporally constant. The Hamiltonian of a ring with $N$ weak linked
SNS-junctions is the following:
\begin{equation}
 H = -E_J \sum_{n=0}^{N-1} \cos(\phi_n - a_n)
\end{equation}
where $E_J$ is the Josephson energy, $\phi_n = \theta({\bf r}_n) - \theta({\bf r}_{n-1})$ is the phase
difference at the junction $n=0,...,N$, and $\theta({\bf r}_n)$ is the phase of
the superconducting island centered at \\
${\bf r}_n = -R \cos(2\pi n/N) {\hat x} + R \sin(2\pi n/N) {\hat y}$ with $R$ the ring
radius.
The magnetic field contribution to the phase difference $a_{n}$ is
the line integral of the vector potential between sites $n$ and
$n-1$,
\begin{equation}
 a_n = \frac{2\pi}{\Phi_0} \int_{{\bf r}_{n-1}}^{{\bf r}_{n}} d{\bf l}.{\bf A}({\bf l}).
\end{equation}
Taking ${\bf B}=B {\bf z}$ and the gauge ${\bf A}= B x {\hat
y}$ for the vector potential ${\bf A}$, and introducing the flux number
through the ring $\Phi/\Phi_0 = \pi R^2 B/\Phi_0$ we have,
\begin{equation}
 a_n =   -\frac{\Phi}{2\Phi_0} \biggl[4\pi/N + \sin(4\pi n/N)- \sin(4\pi(n-1)/N)\biggl]
\end{equation}

We consider the resistive shunted model using a resistive channel
for the normal electron current in parallel with a Josephson
current channel, satisfying Kirchhoff laws for the current
conservation in each node. We inject a current $I$ between
junctions $N-1$ and $0$, and extract it $\delta$ junctions away,
between junctions $\delta-1$ and $\delta$. The resulting set of
dimensionless equations for the currents flowing in the ring is
the following:

\begin{equation}
\dot{\phi}_n = I_{up} - \sin(\phi_n - a_n) , \;\;\;  0 \le n \le
\delta-1
\end{equation}
\begin{equation}
\dot{\phi}_n = I_{up}-I -  \sin(\phi_n - a_n), \;\;\; \delta \le n
\le N-1
\end{equation}
\begin{eqnarray}
I_{up} \equiv (1-\delta/N) I + \frac{1}{N} \sum_{n=0}^{N-1} \sin(\phi_n
- a_n)
\end{eqnarray}

which are $N$ first order differential equations for the time
evolution of the $N$ phase variables $\{\phi_n\}_{n=0}^{N-1}$. Let
us note that each junction interacts with all the others through
$I_{up}(\{\phi_n\}_{n=0}^{N-1})$, the total current in the upper
branch of the circuit, which represents a kind of mean-field
interaction plus a drive. The equations can be easily solved
numerically by using the Runge-Kutta method in order to compute
the instantaneous voltage drop $v$ between source and drain, which
can be expressed as,
\begin{equation}
v =  \sum_{n=0}^{\delta-1} \dot{\phi}_n =  \delta I_{up} - \sum_{n=0}^{\delta-1} \sin(\phi_n - a_n).
\end{equation}
Using this model, the rectified mean dc voltage $V_{dc} = \langle
v \rangle$ is calculated as function of the magnetic field for an
ac-sinusoidal current with different amplitudes $I_{ac}$ in the
low frequency limit. We normalize currents by the single junction
critical current $I_0$ and voltages by $R_N I_0$, with $R_N$ the
resistence of the resistive channel. The results for $N=3$ and
$N=5$, both with the same source-drain distance $\delta=2$, are
shown in Fig.~\ref{fig:JJM}. We can clearly see that both, the
$N=5$ and the $N=3$ devices, can rectify, $i.e.$ $|V_{dc}|>0$, if
$\Phi/\Phi_0 \neq n/2$ with $n$ an integer, and if $I_{ac}$ is
above a critical threshold which is smaller for $N=3$. We can also
observe that the maximum of $|V_{dc}|$ is almost the same in both
cases, although for fixed $\Phi/\Phi_0$,  $V_{dc}$ decays slower
as a function of $I_{ac}$ for $N=3$. More importantly, although
qualitatively, the same oscillations are observed as in the
experiment as a function of vorticity.

The experimental results are measured as a function of temperature
with a constant applied current, while in the model the applied
current is changed, keeping the temperature constant. However, the
effect is similar since both increasing T and $I_{ac}$ have an
analogous influence on the system driving it towards the resistive
state.

In brief, the predicted rectification in this model system is
similar to the effects measured in the Al triangle. It is worth
noticing that from the point of view of the superconducting
condensate our experimental system can be regarded as a multiply
connected structure since the order parameter $\psi$ is maximum at
the vertices and minimum at the sides of the triangle (see sketch
in Fig.~\ref{fig:JJM}). Furthermore, for certain fields and
temperatures $\psi=0$ at the middle of the sides of the triangle
and the system can be actually thought of as a ring-like structure
with SNS-junctions. This scenario is modified by the presence of
contact leads which locally enhance the order parameter. In this
case, sample B having the contacts at the sides can be directly
compared with the five junctions ring whereas the $N=3$ ring
imitates the response of sample C. Indeed, this association can be
further justified by noting in frames I and III (or II and IV) of
Fig.~\ref{fig:JJM} that, for the same $\Phi/\Phi_0$, the $N=3$ and
$N=5$ Josephson circuits have opposite responses (for a fixed
$\delta=2$), as it is also found experimentally by comparing in
the same frames of Fig.~\ref{fig:ratchet} the response of samples
B and C for the same $H$.

In conclusion, we studied the influence of contacts on the
rectification effect in superconducting triangles. We demonstrate
that the sign of the rectification voltage depends on the position
of the current contacts. These findings are in clear agreement
with rectification effects obtained in the framework of the
theoretical model presenting triangle as a micronet of identical
Josephson junctions.

This work was supported by the K.U.Leuven Research Fund
GOA/2004/02 program, the Belgian IAP, the Fund for Scientific
Research -- Flanders (F.W.O.--Vlaanderen), the F.W.O. fellowship
(A.V.S.), and by the ESF "Nanoscience and Engineering in
Superconductivity (NES)" programs.

\newpage

\newpage

\newpage

\newpage

\end{document}